\documentclass[12pt]{iopart}
\usepackage[colorlinks, linkcolor=red, anchorcolor=blue,urlcolor=blue, citecolor=blue]{hyperref}

\newcommand{\beq}{\begin{equation}}
\newcommand{\eeq}{\end{equation}}
\newcommand{\beqa}{\begin{eqnarray}}
\newcommand{\eeqa}{\end{eqnarray}}

\usepackage{bm}
\usepackage{iopams}
\usepackage{graphicx}
\usepackage{cite}
\begin{document}

\title[Fast long-range charge transfer in quantum dot arrays]{Fast long-range charge transfer in quantum dot arrays}

\author{Yue Ban$^1, ^2$, Xi Chen$^3$, Gloria Platero$^1$}
\address{$^1$ Instituto de Ciencia de Materiales de Madrid, CSIC, Sor Juana In\'{e}s de la Cruz 3, E-28049 Madrid, Spain}
\address{$^2$ College of Materials Science and Engineering, Shanghai University, 200444 Shanghai, People's Republic of China}
\address{$^3$ Department of physics, Shanghai University, 200444 Shanghai, People's Republic of China}

\ead{yue.ban@csic.es}
\vspace{10pt}

\begin{abstract}
Charge, spin and quantum states transfer in solid state devices is an important issue in quantum information.
Adiabatic protocols, such as coherent transfer by adiabatic passage have been proposed for the direct charge transfer, also denoted as long range transfer, between the outer dots in a QD array without occupying the intermediate ones. However adiabatic protocols are prone to decoherence.
Aiming to achieve direct charge transfer between the outer dots of a QD array with high fidelity, we propose a protocol to speed up the adiabatic transfer, in order  to increase the fidelity of the proccess. Based on shortcuts of adiabaticity by properly engineering the pulses, fast adiabatic-like direct charge transfer between the outer dots can be obtained. We also discuss the transfer fidelity on the operation time in the presence of dephasing. The proposed protocols for accelerating long range charge and state transfer in a QD array offer a robust mechanism for quantum information transfer, by minimizing decoherence and relaxation processes.
\end{abstract}

%
%
%
%
%

\section{Introduction}
Charge, spin and quantum states transfer in solid state devices is an important issue in nanoelectronics and a very active topic of research. In particular it is relevant for quantum information purposes. With the state-of-the-art technology, surface acoustic waves are able to capture electrons and transport them over long distance \cite{SAW1,SAW2,SAW3,SAW4,SAW5,SAW6,SAW7}. Long-range charge transfer, i.e., direct transfer between the edge dots, mediated by quantum superpositions has been experimentally observed in a triple QD \cite{cotunneling1,cotunneling2,cotunneling3}, while photo-assisted long-range transport has been theoretically investigated \cite{cotunneling4,JAP,cotunneling5} and experimentally observed \cite{cotunneling3}.
\begin{figure}[t]
\begin{center}
\scalebox{0.7}[0.7]{\includegraphics{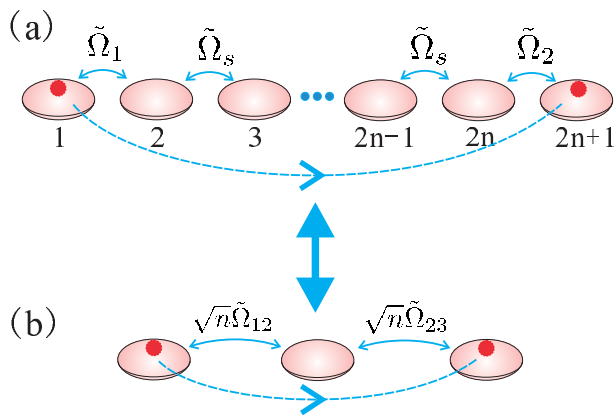}}
\caption{\label{multi-dots} (a) Direct transfer of one electron in an array of $2n+1$ QDs from the $1^{st}$ dot at $t=0$ to the $2n+1^{th}$ dot at $t=t_f$, with the application of the pulses $\tilde{\Omega}_1$, $\tilde{\Omega}_2$ controlling the tunneling between the $1^{st}$ and $2^{nd}$ dots and between the $2n^{th}$ and $2n+1^{th}$ dots, respectively. For $n\geq 2$, straddling pulses $\tilde{\Omega}_s$ are applied between the intermediate dots to suppress the occupation of the internal dots. This scheme can be effectively regarded as a triple QD (b), by modifying  $\tilde{\Omega}_1 \rightarrow \sqrt{n}\tilde{\Omega}_{12}$, $\tilde{\Omega}_2 \rightarrow \sqrt{n}\tilde{\Omega}_{23}$, where $\tilde{\Omega}_{12}$ ($\tilde{\Omega}_{23}$) are the intensities of the pulses between adjacent dots in a triple QD, and where the ``central dot" represents the $2n-1$ ``bulk" QDs with pulses $\tilde\Omega_s$.
}
\end{center}
\end{figure}

In particular, spatial adiabatic passage \cite{reviewBush}, including Coherent Transfer by Adiabatic Passage (CTAP) in a triple QD system \cite{CTAP1}, a variation of Stimulated Raman Adiabatic Passage (STIRAP) \cite{STIRAP1,STIRAP2},
provides an effective scheme to transfer charge directly between the outer dots, avoiding occupation in the middle one \cite{CTAP-period}.
Moreover, such technique has also been extended to the more complicated solid-state quantum computing architectures, such as QD chains with more than three dots \cite{CTAP-donor-chain, CTAP-chain}. In an-all-electrical controlled multi-QD system, CTAP can be realized with straddle coupling between the dots \cite{CTAP1,CTAP-straddle}.
Recent experimental implementations, including scalable gate architecture for up to nine-quantum-dot arrays \cite{petta}, efficient detection and manipulation of charge states in a quintuple QD \cite{Ito} and coherent spin shuttle through a GaAs/AlGaAs quadruple-quantum-dot array \cite{Fujita} motivate us to explore fast protocols for direct transfer between edge dots in quantum dot arrays in order to minimize relaxation and decoherence during the transfer. Our proposed protocol allows to achieve fast long-range charge transfer in long arrays of QDs without exciting the populations in intermediate dots. This will allow to transfer quantum information encoded in a quantum state from one region to a distant one with high fidelity.

Commonly used adiabatic protocols provide slow transfer prone to decoherence, which could reduce the fidelity to some extent. Aiming at reducing the operation time but achieving the adiabatic-like behavior, ``Shortcuts to Adiabaticity" (STA) \cite{reviewSTA}, including counter-diabatic driving (or equivalently transitionless quantum algorithm) \cite{STA-RICE1,STA-RICE2,STA-RICE3,STA-BERRY,STA-CHEN2} and inverse engineering based on the Lewis-Riesenfeld invariant \cite{STA-CHEN1}, provide different ways to speed up the adiabatic passage.
STA has been applied to manipulate single or two interacting spins \cite{STA-BERRY,QiSci}, and to electrically control the spin dynamics of electrons in QDs in the timescale of nanoseconds \cite{STA-single-dot,STA-double-dot}. Combined with STIRAP, STA is also used to speed up the states control with high fidelity up to the non-adiabatic regime in different physical systems, e.g., N-vacancy centers in diamond \cite{STA-STIRAP-NV} or $^{87}$Rb ensembles \cite{STA-STRIAP-atoms}.
Also STIRAP assisted by different implementations of STA, such as counter-diabatic driving in tunnel-coupled quantum wells \cite{STA-SAP}, fast-forward in polyatomic molecules \cite{STA-STIRAP-FF} or non-hermitian shortcuts in coupled optical waveguides \cite{STA-non-hermitian} have already been proposed to control population in three-level systems. To extend the shortcuts to multilevel systems will be of great interest for quantum state transfer in solid state platforms.

In this paper, we report direct and non-adiabatic transfer of one electron between the outer dots in a multi-QD system,
schematically shown in Fig. \ref{multi-dots}. Firstly, we begin with fast charge transfer in a triple QD
based on various STA techniques, including counter-diabatic driving and inverse engineering. Different time-dependent electric pulses are designed to achieve the fast direct charge transfer, by reducing the excitation in the intermediate dot. Detailed comparisons are made to show that counter-diabatic protocols allow fast charge transfer through the triple quantum dot but with finite occupation of the intermediate dot, i.e., they are not suitable for direct transfer between the outer dots.  Inverse engineering however allows to achieve long range charge transfer with high fidelity.
More interestingly, STA for direct charge transfer is further extended to a multi-QD system by inverse engineering. After making the analogy between a triple dot and an array of $2n + 1$ dots, we find the effective pulses, which allow to control the tunneling between the first and the second dots and between the $2n^{th}$ and $2n + 1 ^{th}$ dots. Meanwhile, we obtain their amplitudes, which are proportional to those used in a triple QD.
The fidelity is checked with different operation times in the presence of dephasing by solving the master equation in the Lindblad form. An analytically estimated fidelity is also derived from time-dependent perturbation theory. Combining the numerical and analytical results, we prove that high fidelity can be achieved by shortening
the operation time, up to the order of nanoseconds. 
In the state-of-the-art set up, the charge detection could be easier for the proposed long QD arrays than in a triple QD array as the influence of the outer dots on the detector could be lower than in the three dot case. In addition, the realization of STA in a multi-level quantum system is also extremely useful for other physical systems, like spin transfer in spin chains \cite{CTAP-spin-1} or light propagation in N-coupled waveguides, among others.

\section{Models and Results}
\subsection{Charge Transfer in a triple QD}
\label{triple-dot}

We consider a triple QD in series, where the energy levels are on resonance. The Hamiltonian ($\hbar = 1$ in dimensionless units) reads
\beqa
\label{H0-3level}
H_0(t)= \Omega_{12}(t) c_{1}^\dag c_2 + \Omega_{23}(t) c_{2}^\dag c_3 + h.c.
\eeqa
where two adiabatic pulses of Gaussian shape are applied between adjacent dots controlled by electric gates,
\beqa
\label{pulses-CTAP}
\Omega_{12} = \Omega_0 \exp\left[- \frac{(t-t_f/2 -\tau)^2}{\sigma^2}\right],
\quad
\Omega_{23} = \Omega_0 \exp\left[- \frac{(t-t_f/2 +\tau)^2}{\sigma^2}\right].
\eeqa
The charge state can be transferred adiabatically from $|1\rangle$ to $|3\rangle$ without populating $|2\rangle$ following the CTAP protocol \cite{CTAP1} via the dark state
\beqa
\label{H-3level-dark-state}
|\phi_0\rangle &=& \cos\theta |1\rangle -\sin\theta |3\rangle,
\eeqa
which is an instantaneous eigenstate of $H_0$ with zero energy and where the mixing angle $\tan\theta = \Omega_{12} / \Omega_{23}$ \cite{CTAP1,STIRAP1}.
The fidelity  $F>0.9999$ can be achieved by using the pulses such that the general adiabatic criteria $ \Omega_0 t_f= 100\pi \gg 1$ is fulfilled.
Here we set the pulse intensity about several hundred MHz, $\Omega_0 = 100 \pi$ MHz, and
the operation time for the CTAP protocol $t_f=50$ in units of $2 \pi/\Omega_0$ ($=0.02 \mu s$), corresponding to the timescale of $1 \mu$s.

 Besides the dark state, the other instantaneous eigenstates of $H_0$ , $|\phi_\alpha \rangle = |\phi_\pm \rangle$ have non-zero energies $E_\alpha = E_\pm$ and
\beqa
\label{H-3level-eigenstates}
|\phi_\pm\rangle &=& \frac{1}{\sqrt{2}}(\sin\theta |1\rangle \pm |2\rangle +\cos\theta |3\rangle).
\eeqa
The solution to the time-dependent Schr\"{o}dinger equation of $H_0$ is expressed as $|\Psi(t)\rangle = \sum_j a_j(t) |j\rangle$, in the basis of the on-site states of each dot $|j\rangle$, where the population in each dot is $P_j = |a_j|^2$.
Provided that the initial state is one of the instantaneous eigenstates, the electron will remain in it if the adiabaticity criterion is fulfilled \cite{CTAP1},
\beqa
\label{adiabaticity-criterion}
|E_0-E_\alpha| \gg |\langle \dot{\phi}_0|\phi_\alpha\rangle|.
\eeqa
This can be simplified as $\dot\theta \ll \sqrt{\Omega_{12}^2+\Omega_{23}^2}$ \cite{STIRAP1,STA-transformation}.

\begin{figure}[t]
\begin{center}
\scalebox{0.6}[0.6]{\includegraphics{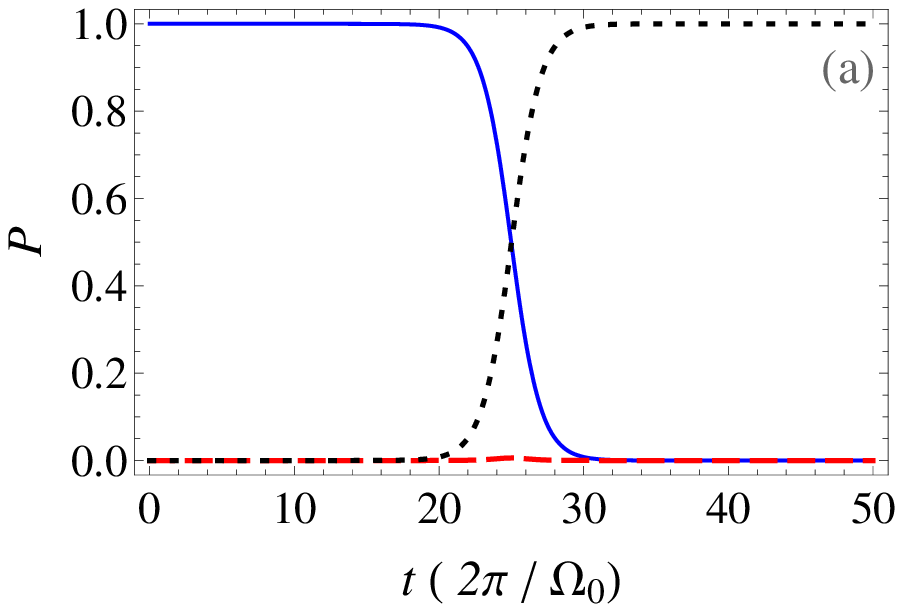}}
\scalebox{0.6}[0.6]{\includegraphics{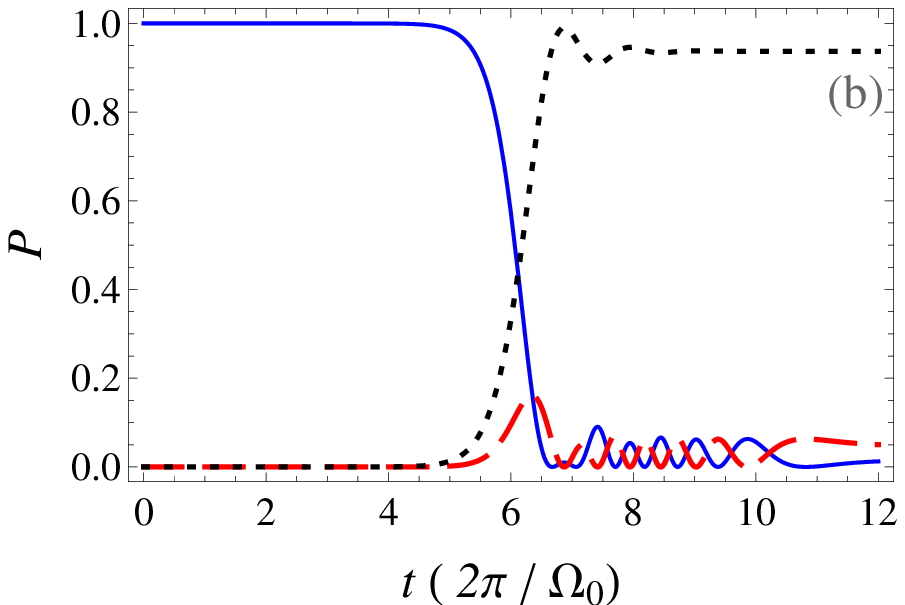}}
\scalebox{0.6}[0.6]{\includegraphics{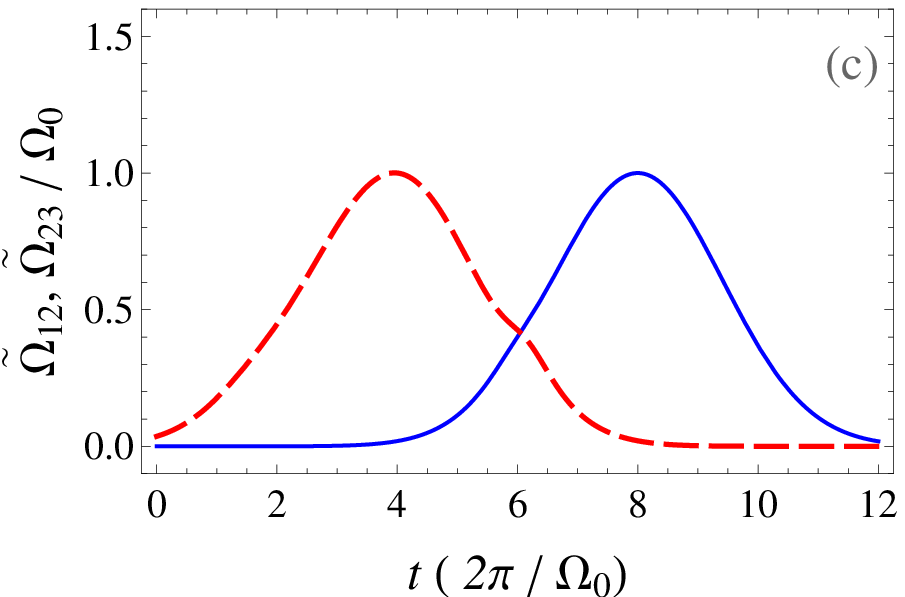}}
\scalebox{0.6}[0.6]{\includegraphics{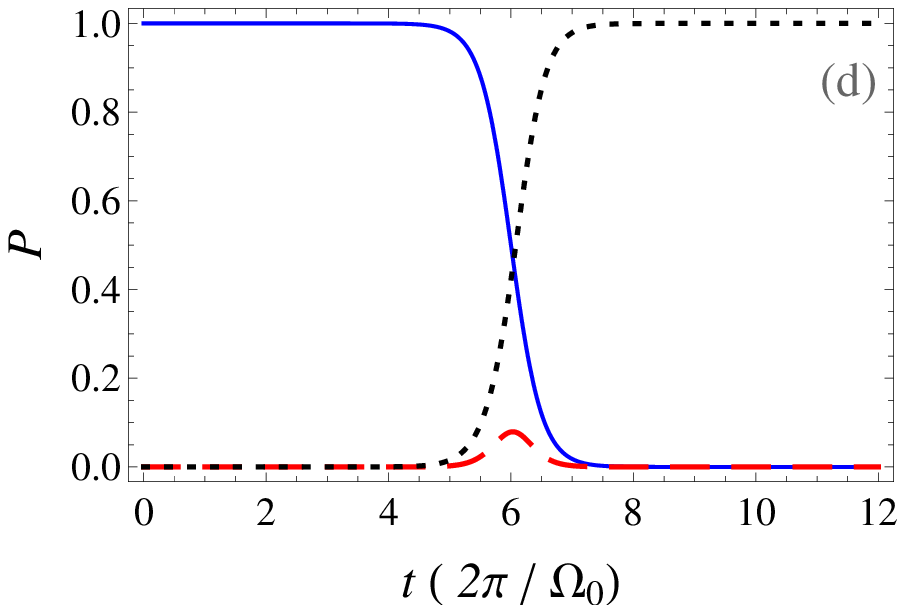}}
\scalebox{0.6}[0.6]{\includegraphics{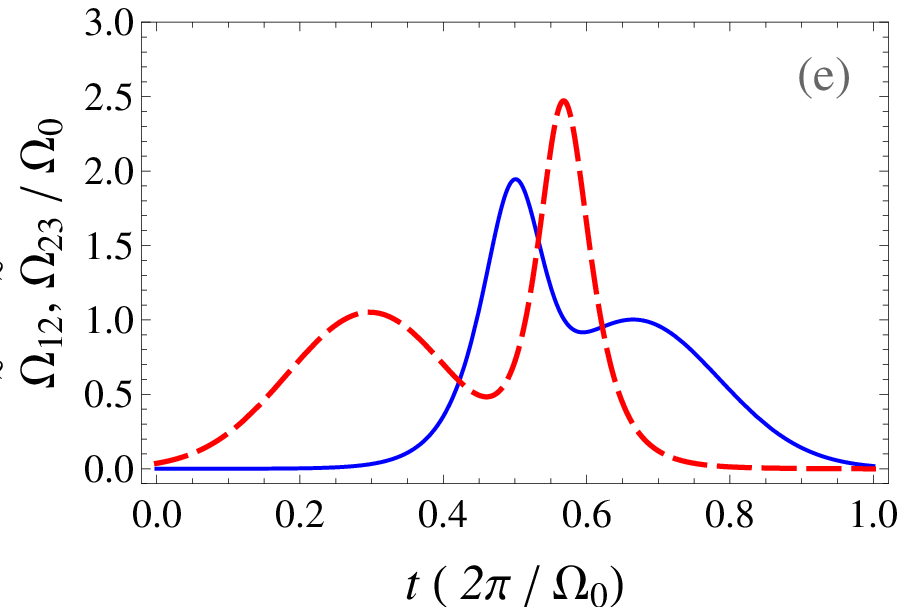}}
\scalebox{0.6}[0.6]{\includegraphics{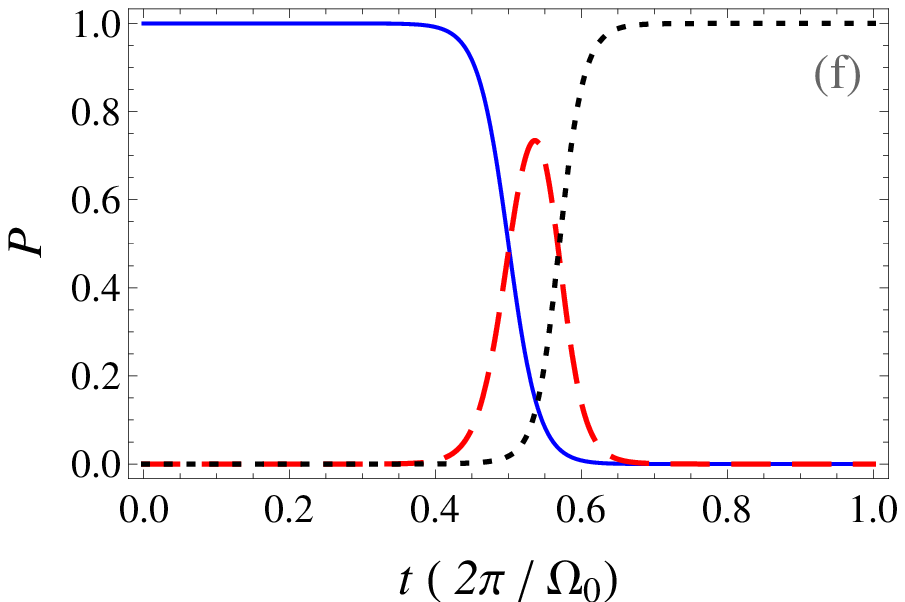}}
\caption{\label{3-levelCD} Population transfer by using CTAP in a triple QD  with $t_f=50$ (a) and $t_f =12$ (b). Shortcuts to adiabatic charge transfer in a triple QD with $t_f=12$ (d) and $t_f=1$ (f), based on counter-diabatic driving. The corresponding designed effective pulses are plotted in (c) and (e). Two pulses $\tilde\Omega_{12}$ (solid, blue), $\tilde\Omega_{23}$ (dashed, red) and population transfer $P_1$ (solid, blue), $P_2$ (dashed, red) and $P_3$ (dotted, black) are depicted. The parameters of the pulses are $\sigma= \tau = t_f/6$. The time is scaled by $2 \pi / \Omega_0$ and $\Omega_0 = 100\pi$MHz.
}
\end{center}
\end{figure}

In order to accelerate the adiabatic transfer, we consider first a triple QD and determine the counter-diabatic Hamiltonian $H_1$ such that
the state evolution is exactly along the $|\phi_0\rangle$, without generating transitions among all the eigenstates of $H_0$.
Using counter-diabatic driving, we can find a time-dependent supplementary interaction \cite{STA-BERRY}
\beqa
\label{H1-3level}
H_1 = i \sum |\partial_t \phi_\alpha \rangle \langle \phi_\alpha| = i \Omega_a c_1^\dag c_3 + h.c.,
\eeqa
with $\Omega_a = \dot\theta$, in order to cancel the diabatic transition.
However, this complementary counter-diabatic term $H_1$ couples the first and the third dots, and it is difficult to implement experimentally.
Thus, we look for physically feasible shortcuts by introducing an appropriate unitary transformation of $H = H_0 +H_1$ \cite{STA-transformation}, which is $\tilde{H} = U^\dag H U - i U^\dag \dot{U}$. Consequently,
\beqa
\label{H-3level}
\label{H3-tilde}
\tilde{H} = \tilde\Omega_{12}c_{1}^\dag c_2 + \tilde\Omega_{23} c_{2}^\dag c_3 + h.c.,
\eeqa
results in the cancellation of this non-adjacent coupling. The adoption of the unitary operator can be done in many ways \cite{STA-transformation}.
For instance, the unitary operator can be \cite{STA-transformation}
\beqa
U = \left(\begin{array}{ccc}
 1 & 0 & 0
\\
0 & \cos\varphi & -i \sin\varphi
\\
 0 & -i \sin\varphi & \cos\varphi
\end{array}
\right).
\eeqa
Consequently, the modified pulses are derived as $\tilde\Omega_{12} = \sqrt{\Omega_{12}^2 + \Omega_a^2}$, $\tilde\Omega_{23} = \Omega_{23} - \dot\varphi$, $\varphi = \arctan (\Omega_a / \Omega_{12})$.
One could also choose other unitary transformation \cite{STA-transformation} or apply an alternative shortcut within dressed-state scheme \cite{ClerkPRL,STA-STIRAP-NV}
to cancel the additional interaction and even suppress the excitation of the intermediate state. Note that $U(0) = U(t_f) =1$ is the necessary boundary condition in order to keep the same dynamics of the system before and after the transformation.
In Fig. \ref{3-levelCD}, we compare the CTAP protocol with the STA based on counter-diabatic driving by taking the parameters  $\sigma= \tau = t_f/6$. Figs. \ref{3-levelCD} (a) and (b) show the charge transfer protocol by CTAP with different operating times $t_f=50$ and $t_f=12$, respectively. We show that at $t_f=50$ long range charge transfer is fully achieved. However as the operating time is reduced to $t_f=12$, the protocol fails and the electron is only partially transferred from the left to the right dot. By contrary, we show in Fig. \ref{3-levelCD} (d) that for the same operating time, $t_f=12$, by considering counter-diabatic driving, the central dot is occupied by approximately a $7\%$, while the electron is fully transfer to the right. Reducing further the operating time, the counter-diabatic protocol allows the full transfer of the charge from left to right but the central dot becomes significantly occupied during the transfer, as shown in Fig. \ref{3-levelCD} (f) for $t_f = 1$.
This is because the intensity of the counter-diabatic term is stronger with shorter $t_f$. Therefore we look for another STA technique as inverse engineering as discussed below.


Inverse engineering is an alternative protocol for STA with high flexibility \cite{STA-3level}. To design the modified pulse directly, we parameterize the solution
as follows,
\beqa
\label{psi-3level}
|\Psi (t)\rangle = \cos \chi \cos\eta |1\rangle - i \sin\eta |2\rangle - \sin\chi \cos\eta |3\rangle,
\eeqa
with the unknown time-dependent parameters $\chi$ and $\eta$. Substituting Eq. (\ref{psi-3level}) back into the time-dependent Schr\"{o}dinger equation for $\tilde{H}$, Eq. (\ref{H-3level}),
we obtain the following auxiliary equations:
\beqa
\label{auxiliary-3level-1}
\dot \chi &=& \tan\eta (\tilde{\Omega}_{12} \sin\chi + \tilde{\Omega}_{23} \cos\chi),
\\
\label{auxiliary-3level-2}
\dot \eta &=& \tilde{\Omega}_{12} \cos\chi - \tilde{\Omega}_{23} \sin\chi.
\eeqa
Once we set the appropriate ansatz for $\chi$ and $\eta$, the pulses $\tilde\Omega_{12}$ and $\tilde\Omega_{23}$  controlling the tunneling can be derived inversely for fast charge transfer. When the transfer becomes adiabatic, that is, $\dot \chi \simeq 0$ and $\dot \eta \simeq 0$, we have $\eta\rightarrow 0$, $\tan \chi \rightarrow \tilde{\Omega}_{12}/ \tilde{\Omega}_{23} $, and the whole state evolution will thus follow the dark state, Eq. \ref{H-3level-dark-state}.

\begin{figure}[t]
\begin{center}
\scalebox{0.5}[0.5]{\includegraphics{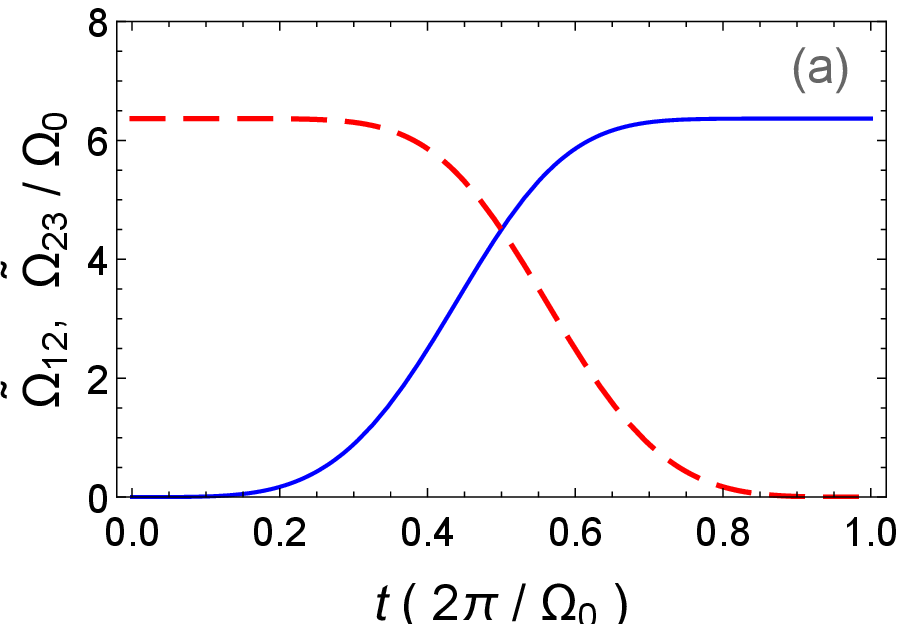}}
\scalebox{0.5}[0.5]{\includegraphics{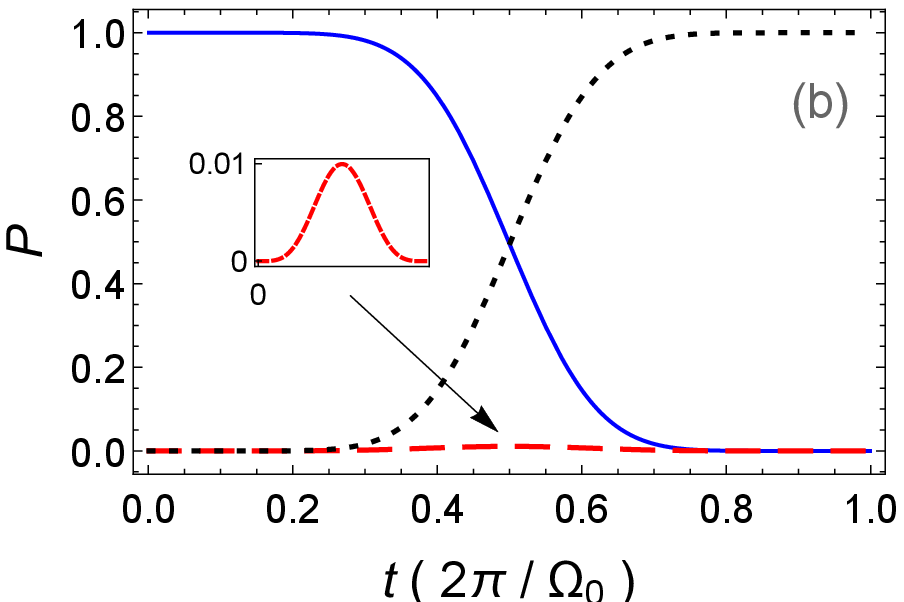}}
\scalebox{0.52}[0.52]{\includegraphics{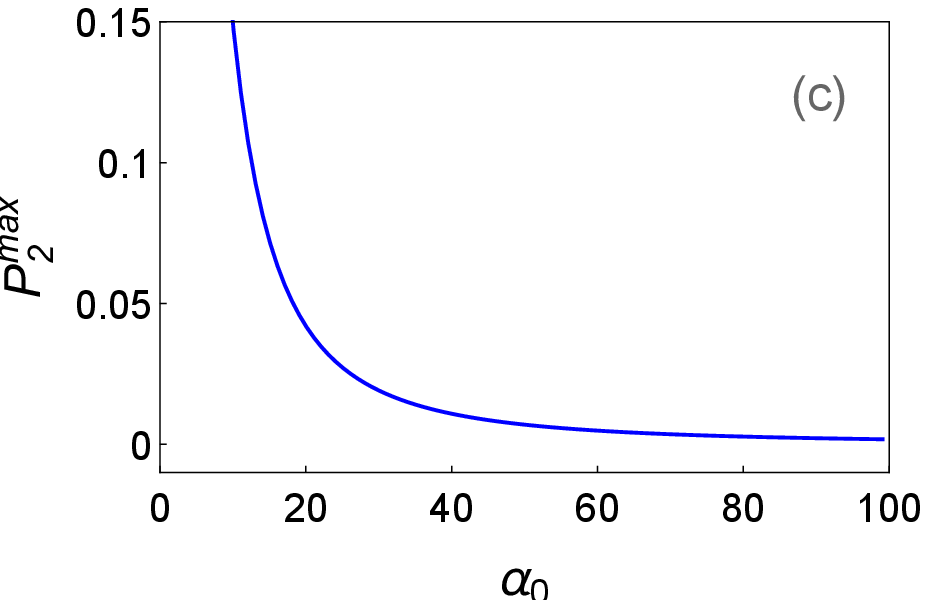}}
\caption{\label{3-levelIE}
Shortcuts to adiabatic charge transfer by means of inverse engineering in a triple QD with $t_f =1$.
(a) Two pulses $\tilde\Omega_{12}$ (solid, blue), $\tilde\Omega_{23}$ (dashed, red) in units of $\Omega_0$ are applied.
(b) The charge occupation of the left ($P_1$, solid blue line), center ($P_2$, dashed red line) and right ($P_3$, black dotted line) dots as a function of time is shown.
Highlighted in the inset, it is depicted that $P_2$ is suppressed below $1$ percent with $\alpha_0 = 40$.
(c) Dependence of $P^{max}_2$ at $t = t_f/2$ and pulse peak on the parameter $\alpha_0$.
Other parameters are the same as those in Fig. \ref{3-levelCD}.
}
\end{center}
\end{figure}

In order to consider STA, the boundary conditions, $\chi(0) = 0$, $\chi(t_f) = \pi/2$, $\eta(0) = 0$ and $\eta(t_f) = 0$ should be imposed for charge transfer from the initial state $|1\rangle$ to final state $| 3 \rangle$ along the solution (\ref{psi-3level}). More boundary conditions: 
$\dot\chi(0) = \ddot{\chi}(0) = \dot\chi(t_f) = \ddot{\chi}(t_f) = 0$ and $\dot\eta(0) = \dot\eta(t_f) = 0$ are required for smooth pulses \cite{STA-3level}, see Eqs. (\ref{auxiliary-3level-1}) and (\ref{auxiliary-3level-2}). To interpolate the functions of $\chi$ and $\eta$,
we adopt the ansatz
\beqa
\label{ansatz-3level}
\chi = \frac{\pi t}{2 t_f} -\frac{1}{3} \sin\left(\frac{2 \pi t}{t_f}\right) + \frac{1}{24} \sin\left(\frac{4 \pi t}{t_f}\right),
\quad
\eta = \arctan{(\dot{\chi}/\alpha_0)},
\eeqa
such that the pulses $\tilde{\Omega}_{12}$ and $\tilde{\Omega}_{23}$
can be calculated from Eqs. (\ref{auxiliary-3level-1}) and (\ref{auxiliary-3level-2}).
In this situation, fast non-adiabatic manipulation inevitably excites the population in the middle dot, which is governed by
\beqa
\label{SE-3}
i \dot{a}_2 = \tilde{\Omega}_{12} a_1 + \tilde{\Omega}_{23} a_3.
\eeqa
In order to reduce the occupation in the central dot and thus to reduce the interaction of the charge in the central dot with the enviroment during the transfer, $\eta$ should be reduced, since the population in the central dot $P_2= \sin^2 \eta$. The parameter $\alpha_0$ provides a degree of freedom to control the amplitude of $\eta$ (see Eq. (\ref{ansatz-3level})). In Fig. \ref{3-levelIE}, we choose $\alpha_0 =40$ as an example. If we compare with CTAP and STA based on counter-diabatic driving (Fig. \ref{3-levelCD}), we observe that the required time to transfer the charge from left to right is strongly reduced.


Therefore, inverse engineering is a robust STA protocol for long range charge transfer in a triple QD array. Fig. \ref{3-levelIE} (a) and (b) depict the designed pulses and the occupation during the transfer, respectively. The occupation of the middle dot is reduced below a $1\%$ by choosing $\alpha_0=40$ for $t_f=1$. For a given $t_f$, reducing the occupation of the middle dot further requires stronger effective pulses by increasing $\alpha_0$." Fig. \ref{3-levelIE} (c) further clarifies the dependence of $P_2^{max}$ at $t=t_f/2$ with $\alpha_0$.

\subsection{Charge transfer through a Multi-QD}

The mechanism of adiabatic charge transfer through a multi-QD system can be realized by means of the so called Straddling Coherent Transfer by Adiabatic Passage (SCTAP) \cite{CTAP1,CTAP-straddle}. This protocol consists of applying adiabatic pulses to the barriers connecting the outer dots with their first neighbor and also to the barriers connecting the intermediate dots. It is valid for chains consisting in an odd number of dots, due to the  spectrum symmetry \cite{CTAP-straddle}. In this system, a dark state is formed such that the states corresponding to the even dots do not participate and only those corresponding to the odd ones and to the outer dots are occupied. The Hamiltonian of an electron in such a system, where the energy levels are on resonance, is written in the form
\beqa
\label{H-mutil-dots}
H_0= \Omega_1 c_1^\dag c_2 + \sum_{1<k<2n} \Omega_s c_k^\dag c_{k+1} + \Omega_{2} c_{2n}^\dag c_{2n+1} + h.c.~~~~~~
\eeqa
where a straddling scheme of internal pulses $\Omega_s $ is considered.
The un-normalized dark state is expressed as \cite{CTAP1}
\beqa
\label{dark-state}
|\phi_0\rangle = \cos\theta |1\rangle - (-1)^n \sin\theta |2n+1\rangle - X\left[\sum_{j=2}^{n} (-1)^{j+1} |2j-1\rangle\right],
\eeqa
where $\tan\theta =\Omega_1/\Omega_2$.
The dots are labelled as $1, 2, ..., n, n+1, ..., 2n, 2n+1$  where $N=2n+1$ is the total number of dots in the array. The hopping rates between neighboring dots $\Omega_{12}$, ..., $\Omega_{n,n+1}$, ..., $\Omega_{2n,2n+1}$, are given by
\beqa
\label{pulses}
\Omega_{1} &=& \Omega_{12} = \Omega_0 \exp\left[-\frac{(t-t_f/2-\tau)^2}{\sigma^2}\right],
\nonumber
\\
\Omega_{2} &=& \Omega_{2n,2n+1} = \Omega_0 \exp\left[-\frac{(t-t_f/2+\tau)^2}{\sigma^2}\right],
\nonumber
\\
\Omega_{s} &=& \Omega_{k,k+1} = \Omega_{s0} \exp\left[-\frac{(t-t_f/2)^2}{2\sigma^2}\right],
\eeqa
where $1<k<2n$, $\Omega_0$ and $\Omega_{s0}$ are the maximal amplitudes of the counter-intuitive pulses and straddled transitions, respectively. In the present QD system, Gaussian-shaped pulses $\Omega_1$ and $\Omega_2$ play the same role as the Pump and Stokes pulses in optical systems, respectively. Meanwhile, the quantum states corresponding to the $2l$ dots, $l=1,2...(N-1)/2$, do not participate in the dark state and therefore remain empty.
If we constrain the parameter \cite{CTAP1}
\beqa
\label{X}
X = \frac{\Omega_1 \Omega_2}{\Omega_s \sqrt{\Omega_1^2+\Omega_2^2}} \ll 1,
\eeqa
i.e., $\Omega_{s0} \gg \Omega_{0}$, the undesirable population in the dots $3^{th}$, $5^{th}$, $...$, $2n-1^{th}$ can be effectively limited,
as it has been experimentally verified by Danzl \textit{et. al} \cite{straddle-experiment} in cold atoms systems.
As a result, by applying $\Omega_1$, $\Omega_2$ pulses and the larger amplitude one $\Omega_s$, population can be transferred from the dot $1^{st}$ to the dot $2n+1^{th}$ directly.

We would like to speed up the adiabatic transfer in five coupled QDs, i.e., $2n+1$ dots where $n=2$.
With the application of the Gaussian-shaped pulses described in Eq. (\ref{pulses}), the charge state can be directly transferred from dot 1 to dot 5 by SCTAP with fidelity $F= |\langle 5 | \Psi(t_f) \rangle|^2 > 0.9999$, via the dark state, when $\Omega_{0} t_f =160 \pi$ and other parameters $\sigma = \tau = t_f/6$
fulfill the adiabatic criteria.
In addition to $E_0=0$ with the corresponding dark state, $|\phi_0\rangle = \left(\cos\theta, 0, X, 0, \sin\theta \right)^T$, four eigenvalues $E_1=-E_2$, $E_3= -E_4$ are symmetric with respect to zero energy. In order to have negligible occupation in the central dot, i.e., the 3rd dot, we adopt $\Omega_{s0} = 10 \Omega_{0}$ to satisfy the condition (\ref{X}).

The analytical expressions for the bright states are more complicated than in the triple QD case. However, under the above constraint (\ref{X}) for $\Omega_1$, $\Omega_2$ and $\Omega_s$, they can be simplified as,
\beqa
\label{four-states}
\nonumber
|\phi_1\rangle = \left(-\frac{\sin\theta}{\sqrt{2}}, \frac{1}{2}, 0, -\frac{1}{2}, \frac{\cos\theta}{\sqrt{2}}\right)^T,
|\phi_2\rangle = \left(-\frac{\sin\theta}{\sqrt{2}}, -\frac{1}{2}, 0, \frac{1}{2}, \frac{\cos\theta}{\sqrt{2}}\right)^T,
\\
|\phi_3\rangle = \left(0, -\frac{1}{2}, \frac{1}{\sqrt{2}}, -\frac{1}{2}, 0\right)^T,
|\phi_4\rangle = \left(0, \frac{1}{2}, \frac{1}{\sqrt{2}}, \frac{1}{2}, 0\right)^T.
\eeqa
By detailed inspection of the Hamiltonian, we find that the counter-diabatic term for the five-QD system becomes
\beqa
\label{H1-5level}
H_1 = i \Omega_a c_1^\dag c_5 + h.c.,
\eeqa
where the additional non-local interaction term $\Omega_a = \dot\theta$ couples the two edge dots. To trace out such non-adjacent interaction in a five-dot system is complicated because it would require to find the appropriate unitary transformation. Therefore we search for mapping our five-QD system to an effective  three-dot one as follows.
\begin{figure}[t]
\begin{center}
\scalebox{0.6}[0.6]{\includegraphics{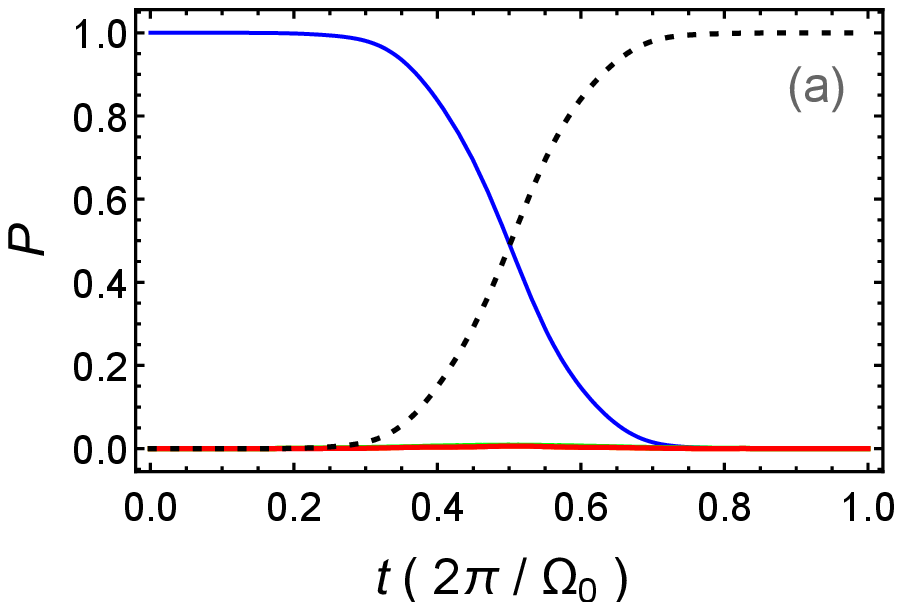}}
\scalebox{0.62}[0.62]{\includegraphics{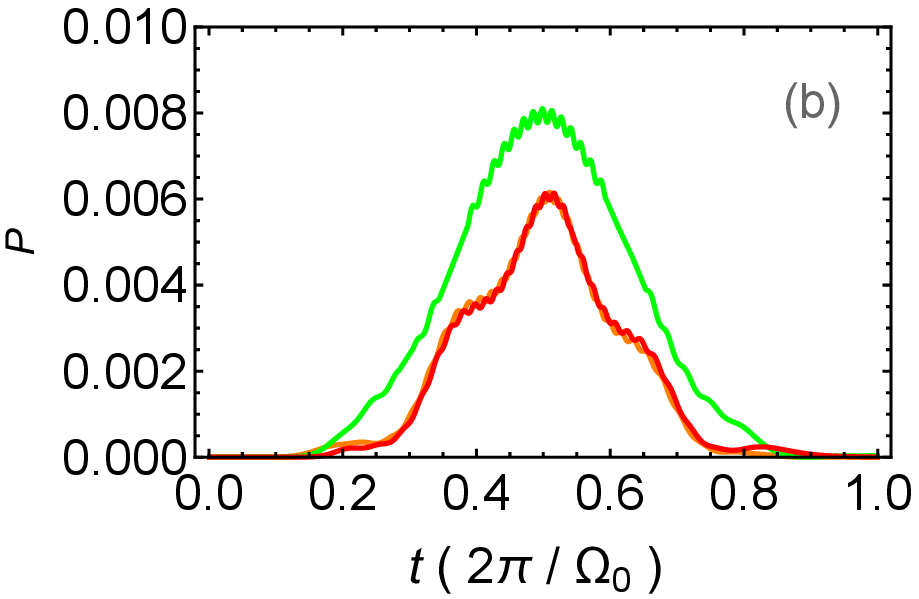}}
\caption{\label{5-level-STA} Speeded-up charge transfer by STA in a five-QD system. (a) Population transfer of all the states in the five dots $P_1$ (solid, blue),  $P_2$ (solid, orange), $P_3$ (solid, green), $P_4$ (solid, red) and $P_5$ (dotted, black), with $t_f=1$. $P_j$ $(j=2, 3, 4)$ are undistinguishable due to their negligible values.
Other parameters are $\tilde{\Omega}_{1} = \sqrt{2} \tilde{\Omega}_{12}$, $\tilde{\Omega}_{2} = \sqrt{2} \tilde{\Omega}_{23}$, $\tilde{\Omega}_{s} = 10 \Omega_s$, where $\tilde{\Omega}_{12}$ and $\tilde{\Omega}_{23}$ are designed by the inverse engineering protocol for the triple dot system, as shown in Fig. \ref{3-levelIE}.
(b) Amplification of $P_2$ (solid, orange), $P_3$ (solid, green) and $P_4$ (solid, red), where the maximum of their amplitudes are less than $1$ percent.
}
\end{center}
\end{figure}
Similarly to the middle dot in a triple QD, population in the $2^{nd}$ and $4^{th}$ dots of a five-QD is excited by fast non-adiabatic manipulation inevitably. We label $b_j$, the probability amplitude of each on-site state $|j\rangle$ for a five-QD to avoid confusion with a triple QD. To reduce the population of  the $2^{nd}$ and $4^{th}$ dots, we make an analogy with the triple QD by keeping $|b_1| \rightarrow |a_1|$, $|b_5| \rightarrow |a_3|$ and by using $\tilde{\Omega}_s$ to achieve ideally $b_3$ equal to zero. Then, we obtain finite but small occupations for the internal even dots. Their corresponding amplitudes fulfill: $|b_2|^2+|b_4|^2 \rightarrow |a_2|^2$. 
Due to the symmetry of the wavefunction $b_2 \rightarrow -b_4$, we find $|b_2-b_4|^2 \rightarrow |b_2|^2 + 2|b_2||b_4| + |b_4|^2 \rightarrow 2|a_2|^2$, and obtain $|b_2-b_4| \rightarrow \sqrt{2}|a_2|$, $i(\dot{b}_2 -\dot{b}_4) \rightarrow \sqrt{2} i \dot{a}_2$. Comparing the relation obtained from the Schr\"{o}dinger equation
\beqa
\label{SE-5}
i (\dot{b}_2 - \dot{b}_4) &=& \tilde{\Omega}_{1} b_1 - \tilde{\Omega}_{2} b_5,
\eeqa
with Eq. (\ref{SE-3}) for a triple QD, the effective pulses for the five-QD system are related to the ones for the triple QD as follows: $\tilde{\Omega}_1 \rightarrow \sqrt{2}\tilde{\Omega}_{12}$ and $\tilde{\Omega}_2 \rightarrow \sqrt{2}\tilde{\Omega}_{23}$. With the strategy discussed above, we find a new Hamiltonian $\tilde{H}$ with modified pulses $\tilde\Omega_{1}$, $\tilde\Omega_{2}$. In order to achieve the suppression of the charge occupation of the central dots, we design $\tilde{\Omega}_{12}$ and $\tilde{\Omega}_{23}$ by inverse engineering as we did for the triple QD (Fig. \ref{3-levelIE}). In Fig. \ref{5-level-STA}, the population transfer through a five-QD array as a function of time is shown. During the operating time $t_f=1$, the charge is fully transferred from the left dot to the right one, with high Fidelity ($F>0.9999$), while the occupancy of the central dots during the transfer remains below $1\%$.

For a multi-dot system with $N=2n+1$ dot, the counter-diabatic term always manifests itself as the interaction between the two outer dots,
\beqa
\label{H1-2n+1level}
H_1 = i \Omega_a c_1^\dag c_{2n+1} + h.c.
\eeqa
We look for a new Hamiltonian $\tilde{H}$ with appropriate $\tilde\Omega_1$ ($\tilde\Omega_2$) controlling the tunneling between the $1^{st}$ ($2n^{th}$) and the $2^{nd}$ ($2n+1^{th}$) dots, which could transfer the charge directly and suppress the occupation in the intermediate dots.
Making analogy with the triple QD, we obtain, for the $2n+1$ QD system, the following expression for the Schr\"{o}dinger equation
\beqa
\label{SE-7}
i \sum\limits_{j=1}^n(-1)^{j+1} \dot{b}_{2j} = \tilde{\Omega}_{1} b_1 + \tilde{\Omega}_2 (-1)^{n+1} b_{2n+1},
\eeqa
equivalent to Eq. (\ref{SE-3}) in a triple QD. Furthermore, in order to transfer the charge directly between the outer dots, following the analogy with the triple dot we impose $|b_1| \rightarrow |a_1|$, $|b_{2n+1}| \rightarrow |a_3|$ and $|b_2|^2 + |b_4|^2 +...+|b_{2n}|^2 \rightarrow |a_2|^2$.
Finally we derive the next general expression for a $2n+1$ QD system: $|\sum_{j=1}^n(-1)^{j+1} {b}_{2j}| \rightarrow \sqrt{n} |a_2|$.
Furthermore, the occupation of the even order dots is suppressed by considering:
$\tilde{\Omega}_{1} \rightarrow \sqrt{n}\tilde{\Omega}_{12}$ and $\tilde{\Omega}_{2} \rightarrow \sqrt{n}\tilde{\Omega}_{23}$, where
$\tilde{\Omega}_{12}$ and $\tilde{\Omega}_{23}$ are obtained in a triple QD by inverse engineering.
The fast and direct charge transfer between the outer dots of a multi-QD system is achieved by manipulating the designed pulses satisfying the above conditions, including Eq. \ref{X}.

\section{Discussion}
\subsection{Reduction of the occupation in the intermediate dots}
Inverse engineering presents a straightforward way to design the desired pulses in order to reduce the occupation of the central dots as discussed in Section \ref{triple-dot}.
In particular, to achieve long range transfer in a triple QD, an adjustable parameter $\alpha_0$ offers a way to suppress the occupation in the middle dot. In a multi-dot system, the population of the $3^{th}, 5^{th}, ..., 2n-1^{th}$ dots is suppressed by increasing the intensity $\tilde\Omega_s$, while that of the even-number dots is lowered by using $\tilde{\Omega}_{1} \rightarrow \sqrt{n}\tilde{\Omega}_{12}$ and $\tilde{\Omega}_{2} \rightarrow \sqrt{n}\tilde{\Omega}_{23}$, where $\tilde{\Omega}_{12}$ and $\tilde{\Omega}_{23}$ are $\alpha_0$-dependent.

Furthermore, we have considered and compared the efficiency of an alternative shortcut technique based on a dressed states framework \cite{ClerkPRL} (not shown in the text). The required intensities of pulses with this technique are larger than those required with the one proposed here. Therefore, with the present protocol, heating effects are minimized. Besides, inverse engineering in a two-level system allows for further optimization by adding one more parameter \cite{QiSci}. Such flexibility provides the freedom to optimize STA with respect to various cost functions, i.e., peak intensity of pulses and fidelity, by combining with optimal control theory \cite{PRLDijon,NJPSherson}.
The optimization procedure for a multi-QD system will be implemented in further work.

\subsection{Relation between $\tilde\Omega_{max}$ and $t_f$}
\begin{figure}[t]
\begin{center}
\scalebox{0.6}[0.6]{\includegraphics{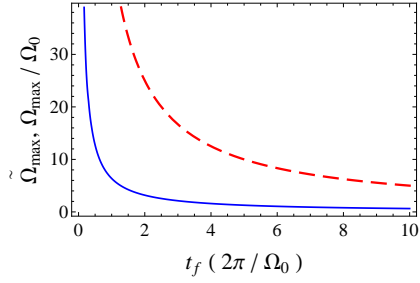}}
\caption{\label{tf-Omegamax} Relation between the pulse peak $\tilde\Omega_{max} (\Omega_{max})$ and $t_f$ by using inverse engineering (solid, blue) and CTAP (dashed, red) for the charge transfer in a triple QD. Other parameters are the same as those in Figs. \ref{3-levelCD} and \ref{3-levelIE}.
}
\end{center}
\end{figure}

In order to show the efficiency of our protocol we compare the behavior of the pulse peak $\tilde\Omega_{max} (\Omega_{max})$ for inverse engineering and CTAP in a triple QD, when the fidelity is above $0.9999$ in both cases. As shown in Fig. \ref{tf-Omegamax}, in a triple QD, inverse engineering is more efficient than CTAP, as the maximal intensities of the designed pulses are smaller for a given $t_f$. Here
we write down the explicit expressions of the shortcut pulses,
\beqa
\label{IE-pulses-3level-a}
\tilde\Omega_{12} &=&  \dot\eta \cos\chi +  \dot\chi  \cot \eta \sin \chi,
\\
\label{IE-pulses-3level-b}
\tilde\Omega_{23} &=& -\dot\eta \sin\chi + \dot\chi  \cot \eta  \cos \chi,
\eeqa
from Eqs.  (\ref{auxiliary-3level-1}) and (\ref{auxiliary-3level-2}). Obviously,
their intensities could become infinite, when $\eta \rightarrow 0, \pi $ and $\cot \eta \rightarrow \infty$.
To avoid the divergence, we introduce the parameter $\alpha_0 = \dot\chi \cot\eta$, Eq. \ref{ansatz-3level}. It results, together with the boundary conditions imposed, in $\tilde\Omega_{max} = \tilde\Omega_{12}(t_f) = \tilde\Omega_{23}(0) = \alpha_0$. Meanwhile,
the maximal population excitation in the second dot is given by
\beq
\label{P2max}
P_2^{max} = \sin^2\eta(t_0)= \frac{\dot{\chi}^2(t_0)}{\dot{\chi}^2(t_0)+\alpha^2_0},
\eeq
with $t_0=t_f/2$, see also Fig. \ref{3-levelIE} (c).
Therefore, one has to increase the intensity of the pulse, $\tilde\Omega_{max}$,
in order to suppress the population excitation in the second dot below $1$ percent, by adjusting the parameter $\alpha_0$.
Moreover, the maximum intensity and operation time satisfy the relation of time-energy uncertainty \cite{STA-double-dot},
as $\dot \chi(t_0) \geq \pi/(2 t_f)$ yields to $\tilde\Omega_{max} t_f \geq \pi / 2\tan\eta(t_0)$.
This rule also holds for longer arrays of QDs, where the value of $P_2^{max}$ (Eq. \ref{P2max}) is amplified by $\sqrt{n}$ times the corresponding value for a triple QD.

\begin{figure}
\begin{center}
\scalebox{0.6}[0.6]{\includegraphics{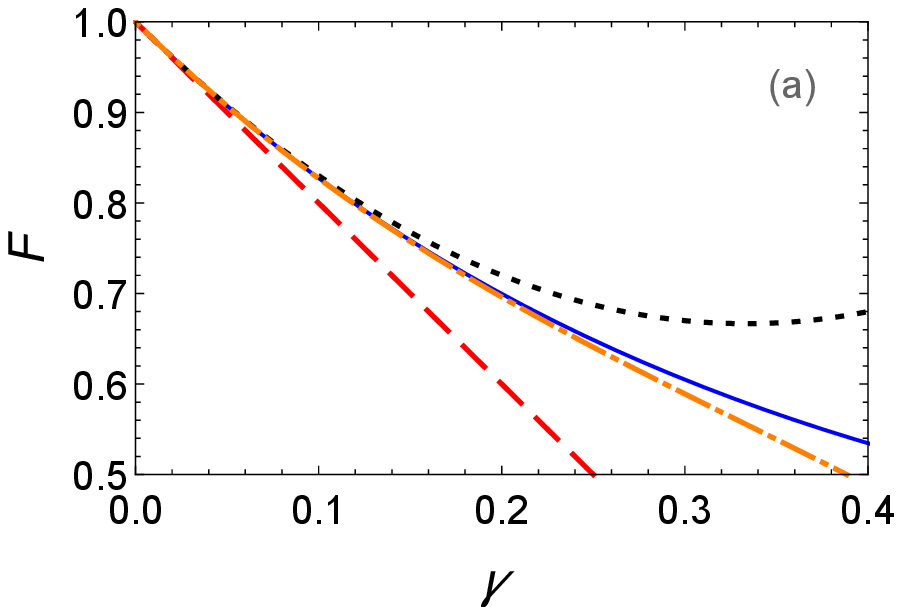}}
\scalebox{0.6}[0.6]{\includegraphics{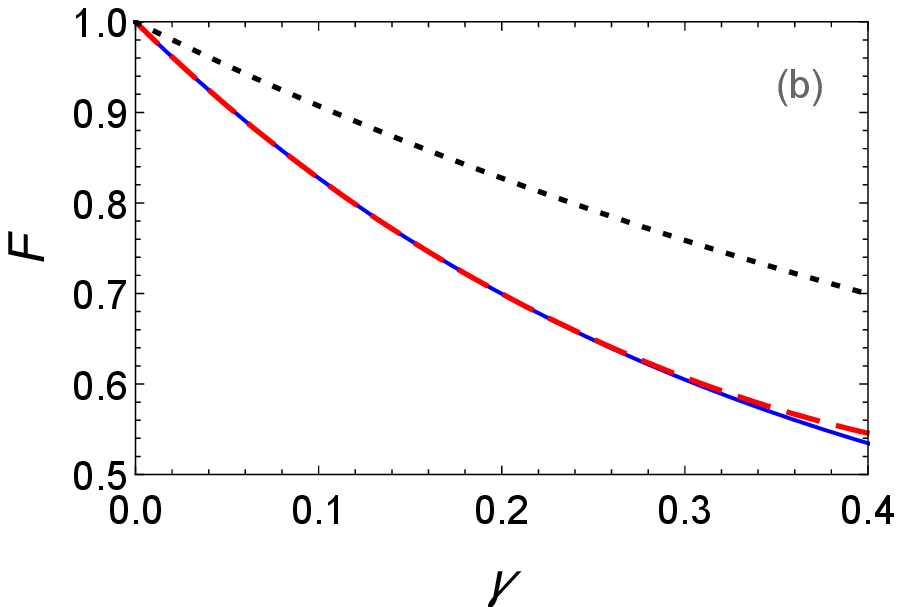}}
\caption{\label{F-gamma} Application of STA (inverse engineering) in a triple QD, (a) fidelity versus dephasing rate $\gamma$ (in units of $5\times10^7 s^{-1}$) obtained from the full numerics (solid, blue), first order time dependent perturbation $F^{(1)}$ (dashed, red), second order $F^{(2)}$ (dotted, black) and  third order $F^{(3)}$ (dot-dashed, orange), where $t_f=1$. (b) Calculation of $F$ up to forth order $F^{(4)}$ (dashed, red). It approaches the numerical result $F$ (solid, blue) for $t_f=1$. Decreasing the operation time up to $t_f = 0.5$ leads to higher Fidelity $F$ (full numerics results; dotted, black).
}
\end{center}
\end{figure}

\subsection{Fidelity vs decoherence}
Decoherence is a key issue for quantum state transfer and manipulation. In the following, we analyze the role of decoherence in order to check the feasibility of our strategy. In QDs systems, there are different sources of decoherence, such as hyperfine interaction or electron-phonon interaction among others. The time evolution of the density matrix is described by means of the Liouville-von Neumann-Lindblad equation \cite{Lindblad-ME} as
\beqa
\label{master-equation}
i \dot\rho = [\tilde{H}, \rho] - \frac{i}{2} \sum_k(L^\dag_k L_k \rho + \rho L^\dag_k L_k - 2 L_k \rho L^\dag_k).
\eeqa
where  $\tilde{H}$ is the Hamiltonian by using STA (For CTAP, $\tilde{H}$ is replaced by $H_0$), $L_k=\sqrt{\gamma} J_k$ are the operators in the system through which decoherence processes are introduced, and $\gamma$ is the dephasing rate in units of $5\times10^7 s^{-1}$.
Here  $J_k$ are the spin operators,
obeying the commutation relation: $[J_k, J_l] = i J_{m} \epsilon_{klm}$ \cite{momentum-operator}, where $\epsilon_{klm}$ is Levi-Civita symbol.

Next, we will consider a triple QD system as an example.  For STA, in the master equation (Eq. (\ref{master-equation})), we take $\tilde{H}$  (Eq. \ref{H3-tilde}) by using inverse engineering. Analogous to the master equation of a two-level system which can be described by a Bloch equation with a three-dimensional vector, the master equation of a three-level system can be recasted into a set of eight equations for the elements of the density matrix. The vector is chosen as
\beqa
\label{vector}
\bm{\xi}(t) = &&[\rho_{11} - \rho_{33}, \frac{1}{\sqrt{3}}(\rho_{11} + \rho_{33} - 2\rho_{22}), \rho_{12}+ \rho_{21}, \rho_{21}+\rho_{12},
\nonumber
\\
&&\rho_{13}+ \rho_{31}, \rho_{31}- \rho_{13}, \rho_{23} + \rho_{32}, \rho_{32} - \rho_{23}]^T,
\eeqa
with the norm $2/\sqrt{3}$. Then the effective time-dependent Schr\"{o}dinger equation can be written as $i \dot{\bm{\xi}}= \mathcal{L}\bm{\xi}$,
where $\mathcal{L} = \mathcal{L}_0 + \mathcal{L}_d$. The unperturbed pulse-controlled part is $ \mathcal{L}_0 = \tilde{\Omega}_{12} J_z + \tilde{\Omega}_{23} J_x$,
whereas the $8 \times 8$ matrix $\mathcal{L}_d$
\beqa
\label{Gamma}
\mathcal{L}_d &=& -i \left(\begin{array}{cccccccc}
3 \gamma & 0 & 0 & 0 & 0 & 0 & 0 & 0
\\
0 & 3 \gamma & 0 & 0 & 0 & 0 & 0 & 0
\\
0 & 0 & \gamma & 0 & 0 & 0 & 0 & 0
\\
0 & 0 & 0 & 3 \gamma & 0 & 0 & 0 & 0
\\
0 & 0 & 0 & 0 & 3 \gamma & 0 & 0 & 0
\\
0 & 0 & 0 & 0 & 0 & \gamma & 0 & 0
\\
0 & 0 & 0 & 0 & 0 & 0 & \gamma & 0
\\
0 & 0 & 0 & 0 & 0 & 0 & 0 & 3 \gamma
\end{array}
\right),
\eeqa
is the one which contributes to dephasing. Here $J_x$ and $J_z$ are the matrices of angular momentum operator for spin $7/2$.
In $\mathcal{L}_d$, the dephasing terms with non-zero values are all diagonal, that demonstrates that decoherence directly acts on all the coordinates of the vector. The factors corresponding to $\xi_3$, $\xi_6$ and $\xi_7$ are $-i \gamma$, different from those of other coordinates of the vector. However, as $\xi_3 = \xi_6 = \xi_7 = 0$ holds during the transfer, this can simplify the calculations in order to derive the fidelity. The fidelity for finding the state $|\psi(t_f)\rangle$ in the third dot at the final time $t_f$ is $F= \rho_{33} (t_f) = |\langle 3 | \psi(t_f) \rangle|^2 = [2 - 3 \xi_1(t_f) + \sqrt{3} \xi_2(t_f)]/6$. To see the exact dependence of the fidelity on $t_f$ and the ansatz adopted by the wavefunction in the presence of the perturbation coming from decoherence, we derive the vector by time-dependent perturbation theory up to high orders, i.e. $\bm\xi = \sum_k \bm\xi^{(k)}, k = 0, 1, 2, ...$, which yields a corresponding expansion of the fidelity $F = \sum_k F^{(k)}, k = 0, 1, 2, ...$. For the zero order expansion without taking any perturbation in consideration, we obtain $\xi^{(0)}_1(t_f) = -1$, $\xi^{(0)}_2(t_f) = 1/ \sqrt{3}$, and the ideal transfer fidelity $F^{(0)} = 1$.
Expanded up to the first order, the non-zero coordinate of the vector at $t_f$ is
\beqa
\label{xi-coordinat}
\nonumber
&&-\xi_1(t_f) + \frac{1}{\sqrt{3}}\xi_2(t_f)= -\xi^{(0)}_1(t_f) + \frac{1}{\sqrt{3}} \xi^{(0)}_2(t_f)
\\
&&-i \int_0^{t_f}dt_1 (-1,\frac{1}{\sqrt{3}},0,0,0,0,0,0)U_0(t_f,t)\mathcal{L}_d \bm{\xi}^{(0)}(t_1),
\eeqa
where $U_0$ is the unperturbed time-evolution operator for the vector, resulting in the fidelity $F = F^{(0)} + F^{(1)} = 1 - 2 \gamma t_f$.
The higher orders ($n \geq 2$) can be further calculated from Dyson series, that is,
\beqa
\label{F-high-order}
\nonumber
F^{(n)} &=& \frac{(-i)^n}{2} \int_0^{t_f} dt_n  \int_0^{t_n} dt_{n-1}...\int_0^{t_2}dt_1  \bm{\xi}(t_f)^T U(t_f-t_n) \mathcal{L}_d(t_n)
\\
&&  U(t_{n} -t_{n-1}) \mathcal{L}_d(t_{n-1})...U(t_2 - t_1) \mathcal{L}_d(t_1) U(t_1) \bm{\xi}(0)
\nonumber
\\
&=& \frac{2}{3} (-3 \gamma t_f)^n \frac{1}{n!}.
\eeqa
From the analytic calculations, we find that $F$ is only related to the dephasing rate $\gamma$ and the operation time $t_f$, whatever ansatz of the wavefuction Eq. (\ref{psi-3level}) we adopt. Fig. \ref{F-gamma} demonstrates the fidelity derived from the numerical calculations and the analytical equations by the time-dependent perturbation theory under different $t_f$. Up to the forth order, the analytical results almost coincide with the numerical ones, with $\gamma$ ranging from $0$ to $0.4$.
As it is expected, for a given $\gamma$, shorter $t_f$ means that the system is less prone to decoherence produced by the environment and therefore the fidelity increases.
Obviously, STA provides an efficient way to yield a higher fidelity under the effects of dephasing by shortening the operation time.

\begin{figure}[t]
\begin{center}
\scalebox{0.3}[0.3]{\includegraphics{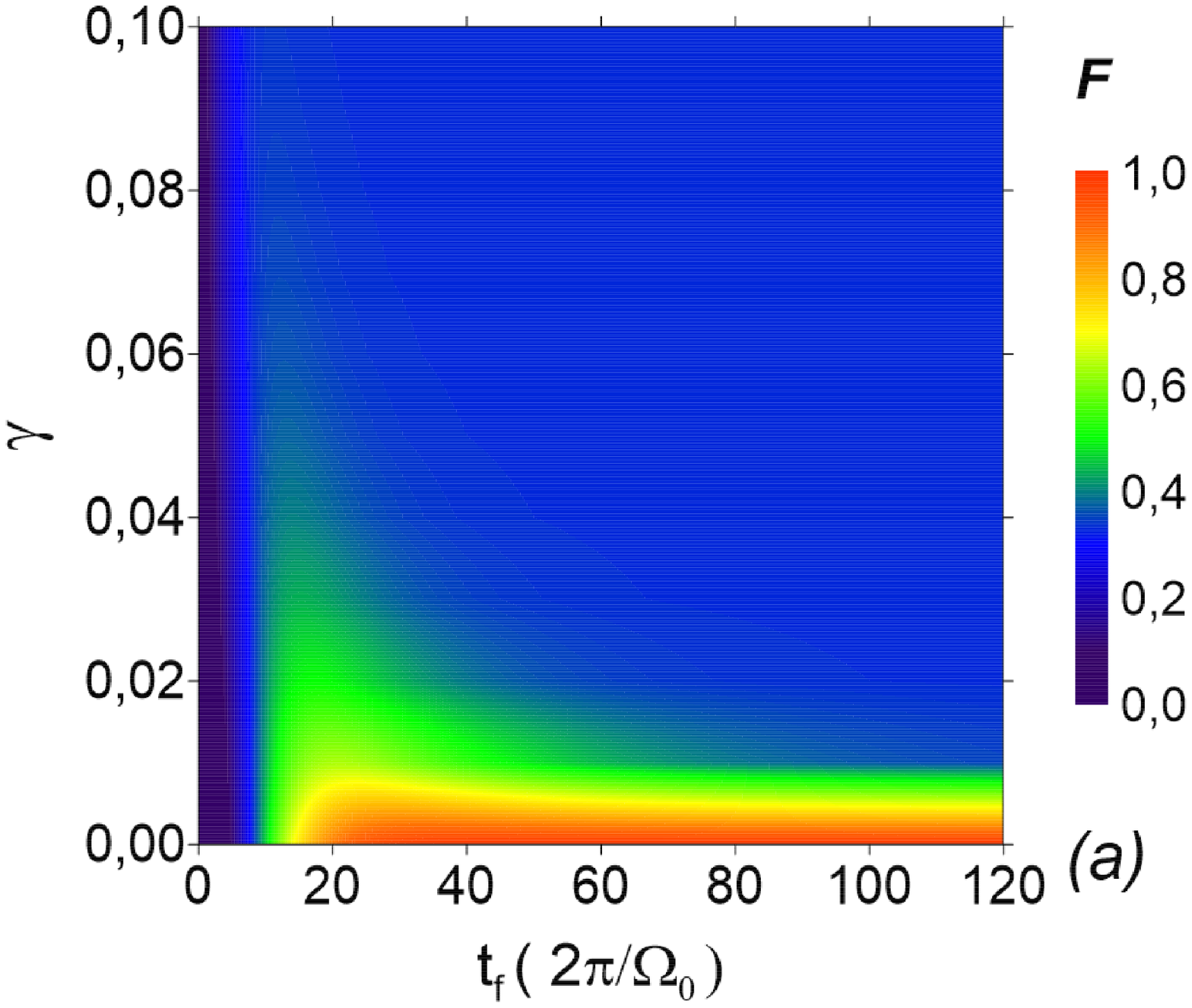}}
\scalebox{0.3}[0.3]{\includegraphics{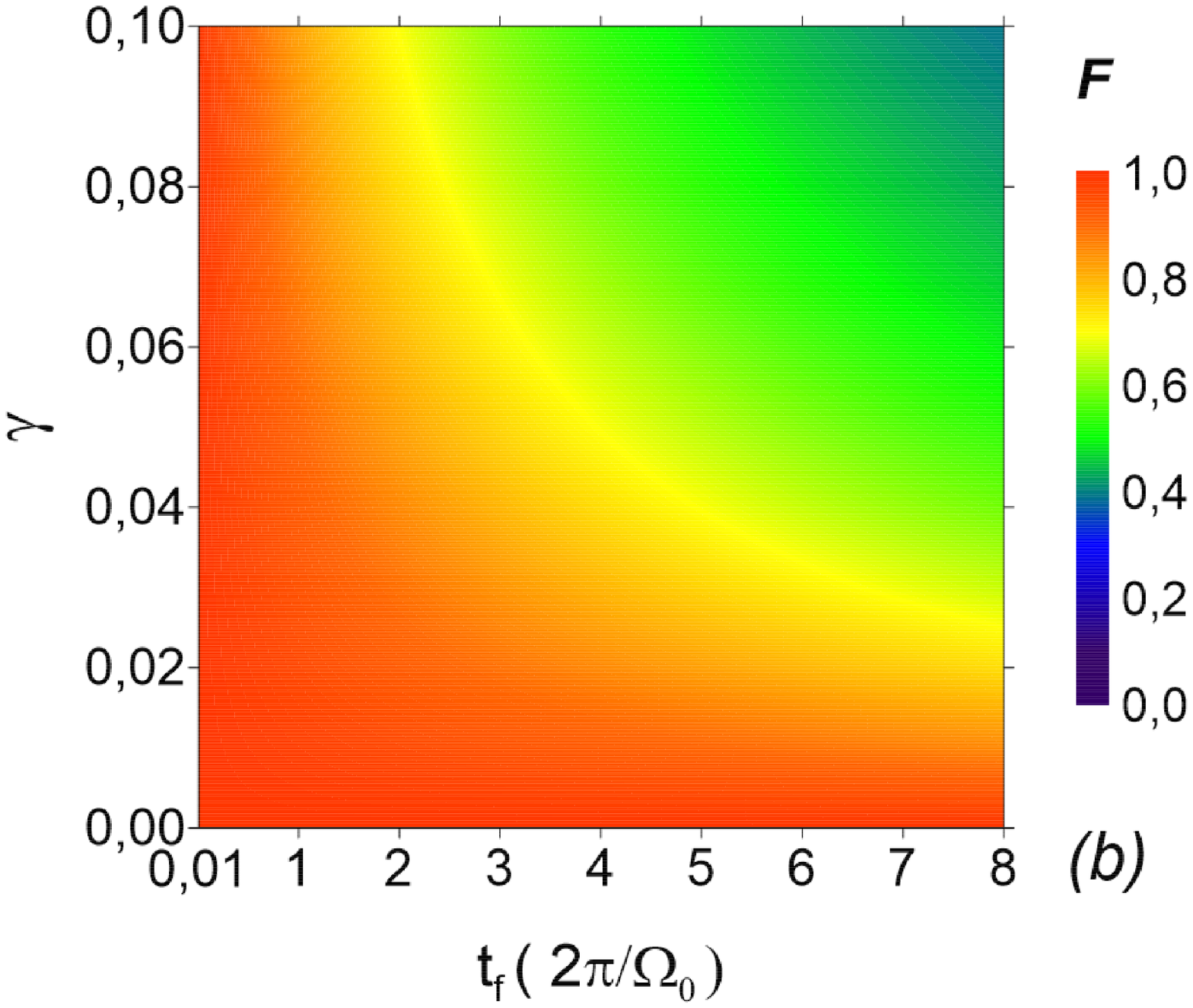}}
\caption{\label{F-gamma-tf} Fidelity $F$ versus dephasing rate $\gamma$ (in units of $5\times10^7 s^{-1}$) and the operation time $t_f$ for a triple QD.
(a): With the application of CTAP, $F$ is high enough just in the adiabatic regime with weak $\gamma$. (b): STA (inverse engineering) extends the high-fidelity regime
to much smaller values of $t_f$, and larger dephasing rates.}
\end{center}
\end{figure}
\begin{figure}
\begin{center}
\scalebox{0.6}[0.6]{\includegraphics{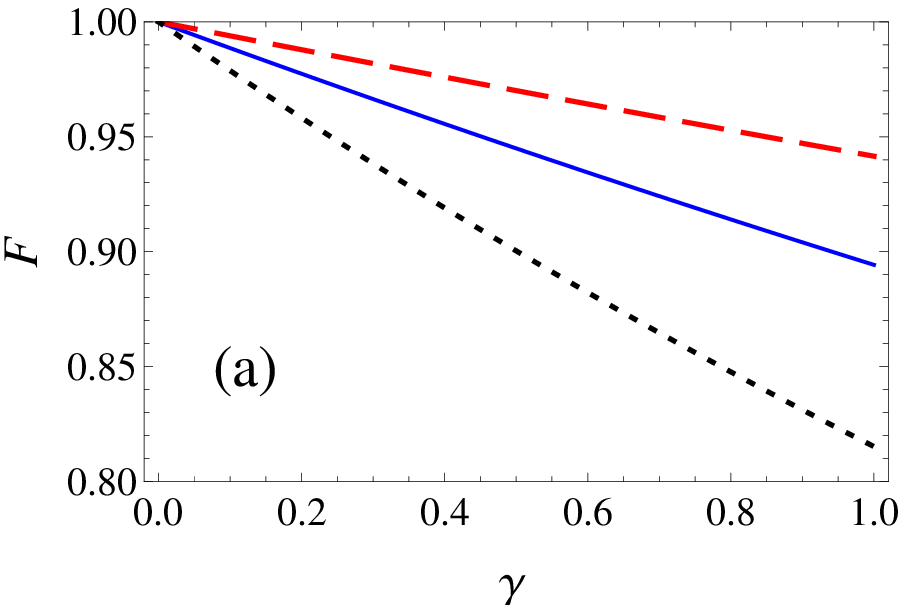}}
\\
\scalebox{0.28}[0.28]{\includegraphics{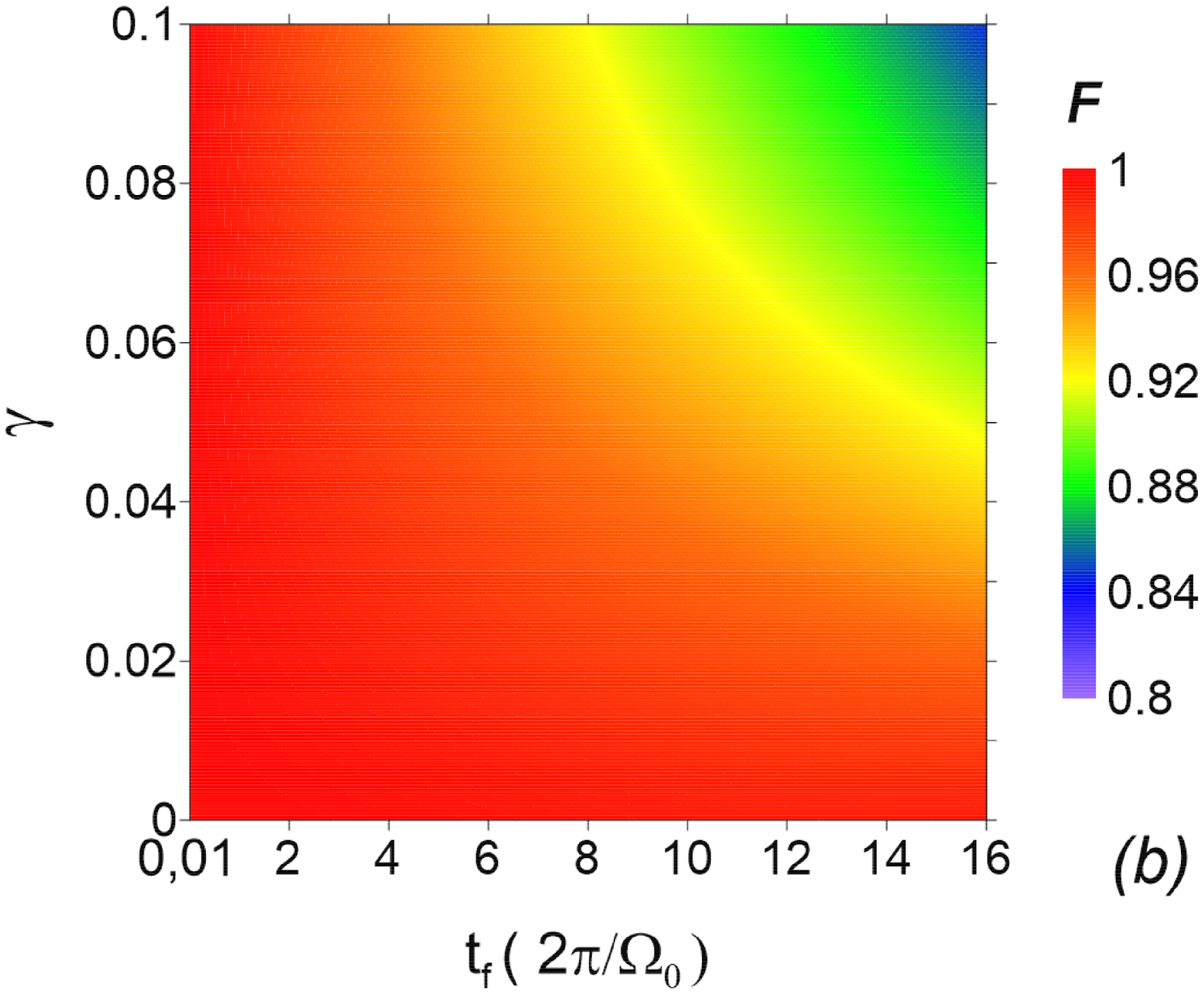}}
\caption{\label{five-F-gamma} (a) Fidelity $F$ of charge transfer in a five-QD system versus dephasing rate $\gamma$ (in units of $5\times 10^7 s^{-1}$) in the presence of different operation time $t_f=0.5$ (red, dashed), $t_f =1$ (solid, blue), and $t_f=2$ (black, dotted). The inverse engineering protocol has been considered. (b) Fidelity $F$ versus dephasing rate $\gamma$ (in units of $5\times 10^7 s^{-1}$) and the operation time $t_f$.}
\end{center}
\end{figure}

Furthermore, we demonstrate in Fig. \ref{F-gamma-tf}, the fidelity $F$ as a function of $t_f$ and $\gamma$ using STA (inverse engineering) and CTAP in a triple QD. By CTAP, only when $t_f>50$ and $\gamma < 0.01$, $\rho_{33}$ approaches to $1$ and the state transfer is fulfilled, due to its vulnerability to decoherence. STA (inverse engineering) extends the regime of direct quantum state transfer to significatively smaller $t_f$. Particularly, when $t_f < 1$, the fidelity remains robust against decoherence. For a multi-dot system with $2n+1$ dots, the master equation can be written as a set of $(2n+1)^2-1$ equations. It is straightforward to analyze the fidelity versus decoherence and its dependence with the operation time, following the procedure developed above for a triple QD. For a five-QD system, we show the numerical results of fidelity as a function of $t_f$ and $\gamma$ by solving the master equation of a five-QD in Fig. \ref{five-F-gamma}. Similar to the case in a triple QD, with shorter $t_f$, fidelity approaches to $1$. As the amplitude of the pulses $\tilde\Omega_1$ and $\tilde\Omega_2$ is $\sqrt{2}$ times of $\tilde\Omega_{12}$ and $\tilde\Omega_{23}$ used in a triple QD, the transfer is more robust against decoherence for the five QD case.

\subsection{Feasibility of Experiments}
With the experimental state of the art, the dynamical occupation of the central dot in a triple QD array can be characterized by means of quantum detection, either by a quantum point contact or a quantum dot detector. Furthermore, recently a quantum-dot device fabricated on an undoped Si/SiGe heterostructure has been demonstrated as a proof of concept for a scalable, linear gate architecture for semiconductor quantum dots \cite{petta}. In their experimental setup, a linear array of nine quantum dots is formed under plunger gates, where tunnel couplings are controlled using barrier gates.  Also the charge state has been detected and manipulated efficiently in a quintuple quantum dot, which provides an important step towards realizing controllable large scale multiple qubits in quantum dot systems \cite{Ito}. In addition, a coherent spin shuttle through a GaAs/AlGaAs quadruple-quantum-dot array has been also achieved \cite{Fujita}. The implementation and high control of semiconductor quantum dot arrays discussed above  motivated the implementation of STA in these systems, in order to improve the speed and stability of direct electron transfer between the edge dots.
Our protocol can be realized in quantum dot arrays of different materials as for instance Si/SiGe or GaAs/AlGaAs. An external time dependent gate voltage in the form of electric pulses is applied between adjacent dots to control the tunneling coupling. In fact, to apply such pulses should be within the state of the art in split-gate structures.
Also, the experimental implementation of the present  proposed protocol becomes even more feasible in longer quantum dot arrays where the detection  of the charge in the central region dots can be achieved without the perturbation of the charge occupation of the outer dots.

For shorter operation times, stronger pulses are needed. However, the intensity of the pulses has to be limited depending of the physical setup in order to avoid strong heating. Therefore, one should find a compromise between pulse intensities and short time operation.

\section{Conclusion}
We report a fast and robust protocol for the long-range charge transfer in large-scale quantum dot arrays without with negligible excitation of the population in the intermediate dots. Initially we consider a triple quantum dot where we explore both counter-diabatic driving and inverse engineering, two STA techniques to speed up the CTAP protocol.
For charge transfer in a multi-dot system, counter-diabatic protocols show that the counter-diabatic term is always located between the first and the last dots. We show that counter-diabatic protocols allow for fast charge transfer but are not suitable protocols for long range, i.e., direct transfer. Inverse engineering protocols however are very efficient ones for direct transfer between not directly coupled sites. They allow to the introduction of a tunable parameter to quench the occupation of the central dot during the transfer. Direct charge transfer without populating the intermediate dots is also achieved in a multi-dot system by analogy to a triple QD.
The relation between the maximal amplitude of the designed pulses and the operation time of our protocol is compared with the one of CTAP. We show that inverse engineering not only speeds up the CTAP protocol but also requires lower pulse intensities than CTAP, reducing possible heating issues. Furthermore, we also check the fidelity against dephasing by solving the master equation in the Lindblad form and prove that higher fidelity is obtained by shortening the operation time by STA. This protocol reveals as a robust mechanism for quantum information transfer, by minimizing decoherence and relaxation processes.
In the case where spin decoherence occurs in a time scale larger than the transfer time obtained by STA (of the order of nanoseconds), all discussed here can be applied to spin transfer. This would be the case of Si quantum dots where hyperfine interaction is negligible.
Spin dephasing time in GaAs quantum dots due to hyperfine interaction is of the order of the time required by STA for transfer. The effect of spin dephasing on the spin transfer protocol will be addressed in a future work.
Furthermore, the proposed transfer protocol is general and also provides an efficient way for optimizing spin qubit manipulation protocols or spin entanglement between distant sites. Moreover, the implementation of the STA protocols for long range transfer of two electron spin states is a promising avenue under investigation.

\section*{Acknowledgements}
\noindent This work is partially supported by the NSFC (Grants No. 61404079 and No. 11474193), the Shuguang
(Grant No. 14SG35), the program of STCSM (Grant No. 18010500400, Grant No. 18ZR1415500), the Program for Professor of Special Appointment (Eastern Scholar) and MINECO via Grant No. MAT2014-58241-P and MAT2017-86717-P. Y. B. also thanks Juan de la Cierva Program by the MINECO (Spain).

\end{document}